\documentclass[american]{tlp}%,draft
\pdfoutput=1 
\usepackage[nodate]{datetime}
\usepackage{babel}
\usepackage{amsmath}
\usepackage{amsfonts,amssymb,stmaryrd}
\usepackage{subfigure}
\usepackage{galois}
\usepackage{mathtools}
\usepackage{tikz}%\usetikzlibrary{snakes,calc,matrix,arrows}

\newtheorem{theorem}{Theorem} % [section]
\newtheorem{definition}{Definition} % [section]
\newtheorem{example}{Example} % [section]

\providecommand*{\coloneq}{\mathrel{\vcentcolon=}}

\newcommand*{\ccp}{\emph{ccp}}
\newcommand*{\tccp}{\emph{tccp}}

\newcommand*{\ra}{\rightarrow}%rightarrow
\providecommand*{\wrt}  {w.r.t.}
\providecommand*{\resp} {respectively}
\providecommand*{\ie}   {i.e.,} 
\providecommand*{\eg}   {e.g.} 

\makeatletter
\providecommand{\ifempty}[3]{\def\@@@temp{#1}\ifx\@@@temp\@empty #2 \else #3\fi}
\makeatother

\providecommand*{\mathoper}[1]{\mathop{\mathit{#1}}\nolimits}
\providecommand*{\llbracket}{[\![}
\providecommand*{\rrbracket}{]\!]}
\providecommand*{\BBrackets}[2][]{\ifempty{#2}{}{\llbracket #2 \rrbracket_{#1}}}
\providecommand*{\syntaxoper}[3]{\mathop{#1}\BBrackets[#3]{#2}}
\providecommand*{\parensmathoper}[2]{\ensuremath{\mathoper{#1}\ifempty{#2}{}{
(#2)}}}

\newcommand*{\prog}{\ensuremath{D}} %% generic program
\newcommand*{\cR}{\prog} %% generic program Alias
\newcommand*{\lub}{\mathit{lub}}
\newcommand*{\glb}{\mathit{glb}}
\newcommand*{\M}{\Rm}
\newcommand*{\Rm}{\mathbb{M}}%dominio sequenze reattive condizionali massimali
\newcommand*{\prefix}{\parensmathoper{prefix}}%op prefissi

\newcommand*{\MGC}{\mathbb{MGC}} %% most general call set
\newcommand*{\dfn}{\coloneq}
\newcommand*{\FUNeq}{\cong} %% variance on tree
\newcommand*{\interpC}[1][]{\mathbb{I}_{#1}} %% interpretations
\newcommand*{\interp}{\interpC} %% interpretations
\newcommand*{\I}[1][\alpha]{\mathcal{I}^{#1}} %% generic interpretation
\newcommand*{\equivClass}[2]{{#1}\big/_{\! #2}}

\newcommand*{\Tp}[3][\alpha]{\syntaxoper{\mathcal{D}^{#1}}{#2}{{#3}}} %%
\newcommand*{\Dd}[2]{\Tp[]{#1}{#2}}%sem dich.
\newcommand*{\ADd}[2]{\Tp{#1}{#2}}%sem dich.
\newcommand*{\Sz}[1][\alpha]{\mathcal{S}^{#1}} %% generic specification
\newcommand*{\Q}[2]{#2 . #1} % D.A -> \Q{A}{D}
\newcommand*{\eval}[3][\alpha]{\syntaxoper{\mathcal{A}^{#1}}{#2}{{#3}}} %%
\newcommand*{\Aa}[2]{\eval[]{#1}{#2}}%sem agente
\newcommand*{\lfp}[1][]{\mathoper{lfp}_{#1}}
\newcommand*{\F}[2][\alpha]{\syntaxoper{\mathcal{F}^{#1}}{#2}{}} %% general lfp
\newcommand*{\prop}[2]{\ifempty{#1}{\mathbin{\odot}}{#1\mathbin{\odot}}#2}%prop
\newcommand*{\parallelseq}[2]{\mathbin{\ifempty{#1}{\dot{\parallel}}{#1\dot{
\parallel} #2}}}%parallel sequenze condizionali

\newcommand*{\Beha}[2][\mathit{ss}]{\syntaxoper{\mathcal{B}^{#1}}{#2}{}} %%
\newcommand*{\BehaQ}[3]{\syntaxoper{\mathcal{B}}{\Q{#2}{#3}}{#1}} %% behavior of
\newcommand*{\ProgEquiv}[3][\mathit{ss}]{#2 \approx_{#1} #3} % program
\newcommand*{\BehaEq}{\ProgEquiv}

\newcommand*{\CSc}[1][c]{#1}
\newcommand*{\CSys}{\mathbf{C}} %constraint system
\newcommand*{\CSdom}{\mathcal{C}} %insieme di constraint
\newcommand*{\CSord}{\preceq}%contrario implicazione
\newcommand*{\CSmerge}{\mathbin{\otimes}}%merge constraints
\newcommand*{\CSjoin}{\mathbin{\oplus}}%merge constraints
\newcommand*{\Var}{\mathit{Var}}%variabili
\newcommand*{\CSimp}{\mathrel{\vdash}}%implicazione
\newcommand*{\CShid}[2][]{\mathop{\exists}\nolimits_{#1}
\ifempty{#2}{}{#2}}%exists

\newcommand*{\askip}{\mathsf{skip}}%agente skip
\newcommand*{\atell}{\parensmathoper{\mathsf{tell}}}%agente tell
\newcommand*{\aask}[1]{\mathop{\mathsf{ask}}\ifempty{#1}{}{(#1)\ra}}
\newcommand*{\asumask}[4][i]{\sum_{#1=1}^{#2}\aask{#3_{#1}}{#4}_{#1}}
\newcommand*{\anow}[3]{\mathop{\mathsf{now}}%agente now
  \ifempty{#1}{}
    {(#1) \mathrel{\mathsf{then}} #2\ifempty{#3}{}{\mathrel{\mathsf{else}} #3}}}
\newcommand*{\ahiding}[3][]{\mathop{\exists^{#1}{#2}} {#3}}
\newcommand*{\aparallel}[2]{#1 \parallel #2}
\newcommand*{\apcall}[2]{#1(#2)}

\newcommand*{\clauseif}	{\ensuremath{\mathbin{\mathord{:}\mathord{-}}}}

\newcommand*{\Agent}{\mathit{Agent}}%Insieme degli agenti

\newcommand*{\ecrs}{\epsilon} %seq. reattiva condizionale vuota
\providecommand*{\pair}[2]{{\langle #1, \, #2 \rangle}}
\newcommand*{\crs}[3]{\ifempty{#1}{}{#1 \ra} \pair{#2}{#3}} %tupla condizionale
\newcommand*{\soddcond}{\mathrel{\rhd}}%soddisfazione di una condizione rispetto
\newcommand*{\ed}{\Box}
\newcommand*{\stutt}[1]{\parensmathoper{\mathit{stutt}}{#1}}

\newcommand*{\cond}[2]{(#1, #2)}
\newcommand*{\uncond}{\mathbin{\otimes_{c}}}
\newcommand*{\Asoddcond}{\mathrel{\tilde{\rhd}}}%soddisfazione di una condizione
\newcommand*{\ACSc}[1][a]{\hat{\CSc[{#1}]}}
\newcommand*{\ACSys}{\hat{\CSys}} %constraint system
\newcommand*{\ACSdom}{\hat{\CSdom}} %insieme di constraint
\newcommand*{\ACSord}{\mathrel{\hat{\CSord}}}
\newcommand*{\ACSmerge}{\mathbin{\hat\CSmerge}}%merge constraints
\newcommand*{\ACSjoin}{\mathbin{\hat\CSjoin}}%merge constraints
\newcommand*{\ACSfalse}{\hat{\CSfalse}}
\newcommand*{\ACStrue}{\hat{\CStrue}}
\newcommand*{\ACShid}[2][]{\mathop{\hat\exists}\nolimits_{#1}
\ifempty{#2}{}{#2}}
\newcommand*{\ACSimp}{\mathrel{\hat\CSimp}}%implicazione

\newcommand*{\ACScDual}[1][a]{\check{\CSc[{#1}]}}
\newcommand*{\ACSysDual}{\check{\CSys}} %constraint system
\newcommand*{\ACSdomDual}{\check{\CSdom}} %insieme di constraint
\newcommand*{\ACSordDual}{\mathrel{\check{\CSord}}}
\newcommand*{\ACSmergeDual}{\mathbin{\check\CSmerge}}%merge constraints
\newcommand*{\ACSjoinDual}{\mathbin{\check\CSjoin}}%merge constraints
\newcommand*{\ACStrueDual}{\check{\CStrue}}
\newcommand*{\ACSfalseDual}{\check{\CSfalse}}
\newcommand*{\ACShidDual}[2][]{\mathop{\check\exists}\nolimits_{#1}
\ifempty{#2}{}{#2}}
\newcommand*{\ACSimpDual}{\mathrel{\check\CSimp}}%implicazione

\newcommand*{\ACSinj}[2]{#1 \mathbin{\hat{\times}} #2}
\newcommand*{\ACSinjDual}[2]{#1 \mathbin{\check{\times}} #2}
\newcommand*{\ACSimppn}{\mathrel{\tilde\vdash}}%implicazione positiva e negativa

\newcommand*{\ACSetap}{\hat{\eta}}
\newcommand*{\ACSetam}{\check{\eta}}
\newcommand*{\ACSetac}{\tilde{\eta}}
\newcommand*{\ACSetapm}{\cond{\ACSetap}{\ACSetam}}

\newcommand*{\Acrs}[4][n]{#2 \ra \pair{#3}{#4}^{#1}} %tupla condizionale
\newcommand*{\Astutt}[2][n]{\parensmathoper{\mathit{stutt}}{#2}^{#1}}
\newcommand*{\AcrsC}[5][n]{\Acrs[#1]{\cond{#2}{#3}}{#4}{#5}} %condizioni
\renewcommand*{\AA}{\mathbb{A}}
\newcommand*{\A}{\AA}
\newcommand*{\leqAA}{\leq} %% order on Abstract Domain
\newcommand*{\leqA}{\leqAA}
\newcommand*{\lubAA}[2]{{\ifempty{#2}{\bigvee #1}{#1 \vee #2}}}      %% l.u.b.
\newcommand*{\lubA}{\lubAA}
\newcommand*{\glbAA}[2]{{\ifempty{#2}{\bigwedge #1}{#1 \wedge #2}}}  %% g.l.b.
\newcommand*{\glbA}{\glbAA}
\newcommand*{\botAA}{\bot} %% bottom of Abstract Domain
\newcommand*{\botA}{\botAA}
\newcommand*{\topAA}{\top} %% top of Abstract Domain
\newcommand*{\topA}{\topAA}
\newcommand*{\latticeA}{\lattice{\A}{\leqA}{\lubA{}{}}{\glbA{}{}}{\botA}{\topA}}

\newcommand*{\AAa}[2]{\eval{#1}{#2}}%sem agente
\newcommand*{\Ii}{\interpC}%dominio interpretazione di un prog.

\newcommand*{\Aprop}[2]{\ifempty{#1}{\mathbin{\tilde{\odot}}}{#1
\mathbin{\tilde{\odot}} #2}}%prop astratta constraints

\newcommand*{\CSfalse}{\mathit{ff}}
\newcommand*{\CStrue}{\mathit{tt}}
\newcommand*{\almeno}{\parensmathoper{\tau^{-}}}%alpha constraint system piu
\newcommand*{\alpiu}{\parensmathoper{\tau^{+}}}%alpha constraint system piu
\newcommand*{\al}{\parensmathoper{\alpha}}
\newcommand*{\ga}{\parensmathoper{\gamma}}

\newcommand*{\leqM}{\sqsubseteq}

\newcommand*{\C}{\M}
\newcommand*{\leqC}{\leqM}
 
\newcommand*{\lubC}{\lubM} 
\newcommand*{\botC}{\botM}

\newcommand*{\topM}{\top}
\newcommand*{\botM}{\bot}
\newcommand*{\glbM}[2]{{\ifempty{#2}{\bigsqcap #1}{#1 \sqcup #2}}}
\newcommand*{\lubM}[2]{{\ifempty{#2}{\bigsqcup #1}{#1 \sqcup #2}}}
\newcommand*{\lattice}[6]{{(#1, \, \mathord{#2}, \, \mathord{#3}, \,
\mathord{#4}, \, \mathord{#5}, \, \mathord{#6})}}
\newcommand*{\latticeM}{\lattice{\M}{\leqM}{\lubM{}{}}{\glbM{}{}}{\botM}{\topM}}
\newcommand*{\latticeC}{\latticeM}

\newcommand*{\depthk}[1][k]{\ensuremath{\mathit{depth}(#1)}}

\begin{document}

\title[Abstract Diagnosis for \tccp]{Abstract Diagnosis for Timed
Concurrent Constraint programs}

\author[M. Comini, L. Titolo and A. Villanueva]
{MARCO COMINI, LAURA TITOLO\\%\thanks{thanks Udine}\\
Dipartimento di Matematica e Informatica\\
University of Udine\\
Via delle Scienze, 206\\
33100 Udine, Italy\\
\email{\{marco.comini,laura.titolo\}@uniud.it}
\and
ALICIA VILLANUEVA\thanks{This work has been partially supported by the
EU (FEDER), the Spanish MICINN under grant
TIN2010-21062-C02-02 and by the Universitat Polit\`ecnica de
Val\`encia under grant PAID-00-10.}\\
Departamento de Sistemas Inform\'aticos y Computaci\'on\\
Universitat Polit\`ecnica de Val\`encia\\
Camino de Vera s/n\\
46022 Valencia, Spain\\
\email{villanue@dsic.upv.es}
}

\pagerange{\pageref{firstpage}--\pageref{lastpage}}
\volume{}
\jdate{January 2011}
\setcounter{page}{1}
\pubyear{2011}

\maketitle

\label{firstpage}

\begin{abstract}
    The \emph{Timed Concurrent Constraint Language} (\tccp\ in short) is a
    concurrent logic language based on the simple but powerful concurrent
    constraint paradigm of Saraswat.
    In this paradigm, the notion of store-as-value is replaced by the
    notion of store-as-constraint, which introduces some differences \wrt\
    other approaches to concurrency.
    
    In this paper, we provide a general framework for the debugging of
    \tccp{} programs.  To this end, we first present a new compact,
    bottom-up semantics for the language that is well suited for debugging
    and verification purposes in the context of reactive systems.
    We also provide an abstract semantics that allows us to effectively
    implement debugging algorithms based on abstract interpretation.

    Given a \tccp{} program and a behavior specification, our debugging
    approach automatically detects whether the program satisfies the
    specification.  This differs from other semi-automatic approaches to
    debugging and avoids the need to provide symptoms in advance.
    We show the efficacy of our approach by introducing two illustrative
    examples.
    We choose a specific abstract domain and show how we can detect that a
    program is erroneous.
\end{abstract}


\section{Introduction}

Finding program bugs is a long-standing problem in software construction.
In the concurrent paradigms, the problem is even worse and the traditional
tracing techniques are almost useless.  There has been a lot of work on
algorithmic debugging \cite{Shapiro82} for declarative languages, which
could be a valid proposal for concurrent paradigms, but little effort has
been done for the particular case of the concurrent constraint paradigm
(\ccp{} in short; \cite{Saraswat93}).  The \ccp\ paradigm is different from
other programming paradigms mainly due to the notion of store-as-constraint
that replaces the classical store-as-valuation model.  In this way, the
languages from this paradigm can easily deal with partial information: an
underlying constraint system handles constraints on system variables.
Within this family, \cite{deBoerGM99} introduced the \emph{Timed Concurrent
Constraint Language} (\tccp{} in short) by adding to the original \ccp{}
model the notion of time and the ability to capture the absence of
information.  With these features, it is possible to specify behaviors
typical of reactive systems such as \emph{timeouts} or \emph{preemption}
actions, but they also make the language non-monotonic.

In this paper, we develop an abstract diagnosis method for \tccp\ using the
ideas of the abstract diagnosis framework for logic programming
\cite{CominiLMV96a}.  This framework, parametric \wrt\ an abstract program
property, is based on the use of an abstract immediate consequence operator
to identify bugs in logic programs.  It can be considered as an extension
of algorithmic debugging since there are instances of the framework that
deliver the same results.  The intuition of the approach is that, given an
abstract specification of the expected behavior of the program, one
automatically detects the errors in the program.  The framework does not
require the determination of symptoms in advance.  In order to achieve an
effective method, abstract interpretation is used to approximate the
semantics, thus results may be less precise than those obtained by using
the concrete semantics.

The approach of abstract diagnosis for logic programming has been applied
to other paradigms \cite{AlpuenteCEFL02,BacciC10absdiag,FalaschiOPV07}.
This research revealed that a key point for the efficacy of the resulting
debugging methodology is the compactness of the concrete semantics.  Thus,
in this proposal, much effort has been devoted to the development of a
compact concrete semantics for the \tccp\ language to start with.
The already existing denotational semantics
are based on capturing the input-output behavior of the system.  However,
since we are in a concurrent (reactive) context, we want to analyze and
debug infinite computations.  Our semantics covers this need and is
suitable to be used not only with debugging techniques but also with other
verification approaches.

Our new (concrete) compact \emph{compositional} semantics is correct and
fully abstract \wrt\ the small-step behavior of \tccp.  It is based on the
evaluation of agents over a denotation for a set of process declarations
$D$, obtained as least fixpoint of a (continuous, monotone) immediate consequence
operator $\Dd{D}{}$.

Thanks to the compactness of this semantics, we can formulate an efficacious
debugging methodology based on abstract interpretation which proceeds by
approximating the $\Dd{D}{}$ operator producing an ``abstract immediate
consequence operator'' $\ADd{D}{}$.  We show that, given the abstract
intended specification $\Sz$ of the semantics of the declarations $\cR$, we
can
check the correctness of $\cR$ by a single application of $\ADd{D}{}$ and
thus, by a static test, we can determine all the process declarations $d\in
D$ which are wrong \wrt\ the considered abstract property.

To our knowledge, in the literature there is only another approach to the
debugging problem of \ccp{} languages, \cite{FalaschiOPV07}, which is also
based on the abstract diagnosis approach of \cite{CominiLMV96a}.  However,
they consider a quite different concurrent constraint language without
non-monotonic features, which we consider essential to model behaviors of
reactive systems.

\section{The Timed Concurrent Constraint language}

The \tccp{} language is particularly suitable to specify both reactive and
time critical systems.  As the other languages of the \ccp{} paradigm
\cite{Saraswat93}, it is parametric \wrt\ a cylindric constraint system.
The constraint system handles the data information of the program in terms
of constraints.
In \tccp{}, the computation progresses as the concurrent and asynchronous
activity of several agents that can (monotonically) accumulate information
in a \emph{store}, or query some information from that store.
Briefly, a cylindric constraint system\footnote{See
\cite{deBoerGM99,Saraswat93} for more details on cylindric constraint
systems.} $\CSys=\langle \CSdom,\CSord, \CSmerge, \CSjoin, \CStrue, \CSfalse,
\Var, \exists \rangle$ is composed of a set of finite constraints $\CSdom$
ordered by $\CSord$, where $\CSjoin$ and $\CSmerge$ are the $\glb$ and
$\lub$, respectively.  $\CStrue$ is the smallest constraint whereas $\CSfalse$
is the largest one.
We often use the inverse order $\vdash$ (called \emph{entailment}) instead of
$\CSord$ over constraints. $\Var$ is a denumerable set of variables and
$\exists$ existentially quantifies variables over constraints (the so called
cylindric operator).

    Given a cylindric constraint system $\CSys$ and a set of process
    symbols $\Pi$, the syntax of agents is given by the following grammar:
    \begin{equation*}
        A::=\askip \mid \atell{c} \mid \asumask{n}{c}{A} \mid
        \anow{}{}{}\, c\, \mathsf{then}\, A_{1}\, \mathsf{else} \, A_{2} \mid
A_{1}\parallel A_{2} \mid
        \ahiding{x}{A} \mid \apcall{p}{\vec{x}}
    \end{equation*}
    where $c$ and $c_{i}$ are finite constraints in $\CSdom$, $p\in\Pi$,
    $x\in \Var$ and $\vec{x}$ is a list of variables $x_1,\ldots,x_n$
    with $1\leq i\leq n$, $x_i\in \Var$.  A \tccp{} program $P$ is an
    object of the form $\Q{A_{0}}{D}$, where $A_{0}$ is an agent, called
    initial agent, and $D$ is a set of process declarations of the form
    $\apcall{p}{\vec{x}} \clauseif A$ (for some agent $A$).

The notion of time is introduced by defining a discrete and global clock: it is
assumed that the $\aask{}$ and $\atell{}$ agents take one time-unit to
be executed.
For the operational semantics of the language, the reader can consult
\cite{deBoerGM99}.  Intuitively, the $\askip$ agent represents the
successful termination of the agent computation.  The $\atell{c}$ agent
adds the constraint $c$ to the current store and stops. It takes one
time-unit, thus the constraint $c$ is visible to other agents from the following
time instant. The store is updated by means of the $\CSmerge$ operator of
the constraint system.  The choice agent $\asumask{n}{c}{A}$ consults the
store and non-deterministically executes (at the following time instant)
one of the agents $A_i$ whose corresponding guard $c_i$ holds in the
current store; otherwise, if no guard is satisfied by the store, the agent
suspends.  The agent $\anow{}{}{}\,c\,\mathsf{then}\,A\,\mathsf{else}\,B$
behaves in the current time instant
 like $A$ (\resp\ $B$) if $c$
is (\resp\ is not) satisfied by the store. The satisfaction is checked
by using the $\CSimp$ operator of the constraint system.
Note that this agent can process
negative information: it can capture when some information is not present
in the store since the agent $B$ is executed both when $\neg c$ is
satisfied, but also when neither $c$ nor $\neg c$ are satisfied.
$A\parallel B$ models the parallel composition of $A$ and
$B$ in terms of maximal parallelism (in contrast to the interleaving
approach of \ccp{}), \ie{} all the enabled agents of $A$ and $B$ are
executed at the same time.
The agent $\ahiding{x}{A}$ is used to 
make variable $x$ local to %the agent UNDERFULL
$A$.  To this end, it uses the $\exists$ operator of the constraint system.
Finally, the agent $p(\vec{x})$ takes from $D$ a declaration of the form
$p(\vec{x}) \clauseif A$ and executes $A$ at the following time instant.
For the sake of simplicity, we assume that the set $D$ of declarations is
closed \wrt\ parameter names.

\section{Modeling the small-step operational behavior of \tccp}\label{sec:Sem}

In this section, we introduce a denotational semantics that models the
small-step behavior of \tccp{}.  Due to space limitations, in this paper we
show %only UNDERFULL
the 
concrete semantics and the most relevant aspects of the abstract one.
The missing definitions, as well as the proofs of all the
results, can be found in \cite{CominiTV11sem}.

Let us formalize the notion of behavior for a set $D$ of process
declarations.
It collects all the small-step computations associated to  $D$ as the
set
of (all the prefixes of) the sequences of computational steps, for
all possible initial agents and stores.
\begin{definition}[Small-step behavior of declarations]\label{def:ssBeha}  
    Let $D$ be a set of declarations, $\Agent$ the set of possible
    agents, and $\rightarrow$ the transition relation given by the operational
    semantics in \cite{deBoerGM99}. 
    The small-step behavior of $D$ is
    defined as follows:
    \begin{equation*}
        \Beha{D} \coloneq \bigcup_{\forall c\in\CSdom, \forall A\in\Agent}
        \BehaQ{c}{A}{D}
    \end{equation*}
    where
    $\BehaQ{c}{A}{D} \coloneq \{c\cdot c_1\cdot\dots\cdot c_{n}\mid \langle
    A,c\rangle\ra\langle A_1,c_1\rangle\ra\dots\ra\langle
    A_{n},c_{n}\rangle\}\cup\{\ecrs\}$.
    We denote by $\BehaEq{}{}$ the equivalence relation between
    declarations induced by $\Beha{}$, namely $\BehaEq{D_1}{D_2}
    \Leftrightarrow \Beha{D_1} = \Beha{D_2}$.
\end{definition}
The
pair $\langle A_i,c_i\rangle$ denotes a configuration where $A_i$ is the agent
to be executed, and $c_i$ the store at that computation step.
Thus, the small-step behavior is the set of sequences of stores that are
computed by the operational semantics of the language. 

There are many languages where a compact compositional semantics has been
founded on collecting the possible traces for the weakest store, since all
traces relative to any other initial store can be derived by instance of
the formers.  
In \tccp{}, this %approach 
does not work since the language is not
monotonic: if we have all traces for an agent $A$ starting from an initial
store $c$ and we execute $A$ with a more instantiated initial store $d$,
then new traces, not instances of the formers, can appear.%\footnote{See

Furthermore, note that, since we are interested in a bottom-up approach, we
cannot work assuming that we know the initial store.  However, when we %have
define the semantics of a conditional or choice agent where some guard
must be checked, we should consider different
execution branches depending on the guard satisfiability.
To deal with all these particular features, 
our idea is that of associating conditions to
computation steps, and to collect all possible minimal hypothetical
computations.

\subsection{The semantic domain}

In \cite{deBoerGM99}, reactive sequences are used as semantic domain for
the top-down semantics. These sequences
are composed of a pair of stores $\langle c,c'\rangle$ for each time
instant
meaning that, given the store $c$, the program produces in one time
instant the store $c'$.  The store is monotonic, thus $c'$ always contains
more (or equal) 
information than $c$. 

As we have explained before, this information is not enough for a bottom-up
approach.\footnote{In a top-down approach, the (initial)
current store is propagated, thus decisions regarding the satisfaction or
not of a given condition can be taken immediately.}
Our idea is to enrich the reactive sequence notion so that we keep information
about the essential conditions that the store must satisfy in order to make
the program proceed.
We define a \emph{condition} $\eta$ as a pair $\eta=(\eta^{+},\eta^{-})$ where
$\eta^{+} \in \CSdom$ (\resp\ $\eta^{-} \in \wp(\CSdom)$) is called positive
(\resp\ negative) component.
A condition is said to be \emph{inconsistent} when its positive component
is $\CSfalse$ or when it entails any constraint in the negative component.
Given %a condition $\eta=(\eta^{+},\eta^{-})$ and 
a store $c \in \CSdom$, we say
that $c$ satisfies
$\eta$ (written $c\rhd \eta$) when $c$ entails $\eta^{+}$,
$\eta^{+}\neq \CSfalse$ and $c$ does not entail
any constraint from $\eta^{-}$. 
An inconsistent condition is satisfied by no store, while the pair
$(\CStrue,\emptyset)$ is satisfied by any store.

A \emph{conditional reactive sequence} is a sequence of \emph{conditional
tuples},
which can be of two forms: (i)
a triple $\crs{\eta}{a}{b}$ that is used to represent a
computational step, \ie{} 
the global store $a$ becomes $b$ at the next time instant only if 
$a\soddcond \eta$, or (ii)
a construct $\stutt{C}$ that models the suspension of the computation due
to an $\aask{}$ agent, \ie{} it represents the fact that there is no guard
in $C$ (the guards of the choice agent) entailed by the current store.  We
need this construct to distinguish a suspended computation from an infinite
loop that does not modify the store.

Our denotations are composed of \emph{conditional reactive sequences}:
\begin{definition}[Conditional reactive sequence]
    A conditional reactive sequence is a sequence of conditional tuples of
    the form $t_1\dots t_n\dots$, maybe ended with $\ed$, such that: for
    each $t_i = \crs{\eta_i}{a_i}{b_i}$, $b_i\CSimp a_i$ for $i\geq 1$, and
    for each $t_j = \crs{\eta_j}{a_j}{b_j}$ such that $j>i$, $a_j \CSimp
    b_i$.  The empty sequence is denoted with $\ecrs$.  $s_1\cdot s_2$
    denotes the concatenation of two conditional reactive sequences $s_1$,
    $s_2$.
\end{definition}

A set of conditional reactive sequences is \emph{maximal} if none of its
sequences is the prefix of another.  By $\M$ we denote the domain of sets
of maximal conditional reactive sequences, whose order is induced from its
prefix closure, namely $R_{1}\leqC R_{2} \Leftrightarrow
\prefix{R_{1}} \subseteq \prefix{R_{2}}$.
$\latticeC$ is a complete lattice.

\subsection{Semantics Evaluation Function for Agents}

In order to associate a denotation to a set of process declarations, we need
first to define the semantics for agents. 
Let us now introduce the notion of interpretation.
\begin{definition}[Interpretations]\label{def:interp}
    Let $\MGC \dfn \{ p(\vec{x}) \mid p\in \Pi$, $\vec{x}$ are distinct
    variables$\,\}$ be the set of \emph{most general calls}.
    An \emph{interpretation} is a function $\MGC \to \C$ modulo
    variance\footnote{\ie\ a family of elements of $\C$, indexed by $\MGC$,
    modulo variance.}.
    Two functions $I,J :\MGC \to \C$ are \emph{variants}, denoted by $I
    \FUNeq J$, if for each $\pi \in \MGC$ there exists a variable renaming
    $\rho$ such that $(I\pi)\rho =J(\pi\rho)$.
    The semantic domain $\interp$ %_{\Symbols}$ 
    is the set of all interpretations ordered by the point-wise extension
    of $\leqC$.
\end{definition}    
The application of an interpretation $\I[]$ to a most general call $\pi$,
denoted by $\I[](\pi)$, is the application $I(\pi)$ of any representative
$I$ of $\I[]$ which is defined exactly on $\pi$.  For example, if $\I[] =
\equivClass{ ( \lambda \varphi(x,y).\,
\{{\crs{(\CStrue,\emptyset)}{\CStrue}{x=y}}\} ) }{\FUNeq}$ then
$\I[](\varphi(u,v)) = \{{\crs{(\CStrue,\emptyset)}{\CStrue}{u=v}}\}$.

The technical core of our semantics definition is the agent semantics
evaluation function which, given an agent and an interpretation, builds the
maximal conditional reactive sequences of the agent.
\begin{definition}[Agents Semantics]\label{def:semAg}
    Given an agent $A$ and an interpretation $\I[]$, the semantics
    $\Aa{A}{\I[]}$ is defined by structural induction: 
    \par\vskip -3ex {\small
    \begin{align}
        %skip
        & \Aa{\askip}{\I[]} = \{\ed\} \notag \\ %\label{eq:Aa_skip}
        %tell
        & \Aa{\atell{c}}{\I[]} = \{\crs{(\CStrue,\emptyset)}{\CStrue}{c}\cdot\ed\}
        \label{eq:Aa_tell} \\
        %ask   
        & \Aa{\textstyle\asumask{n}{c}{A}}{\I[]} = \lubC {
        \lubC{}{}_{i=1}^{n} \{\crs{(c_{i},\emptyset)}{c_{i}}{ c_{i}}\cdot
        (\prop{c_{i}}{s}) \mid s \in \Aa{ A_{i}}{\I[]}\} } {{}}
        \notag\\*&\qquad \lubC { \{ \stutt{\cup_{i=1}^{n}c_{i}}\cdot s \mid
        s\in \Aa{\textstyle\asumask{n}{c}{A}}{\I[]},\forall
        i\in[1,n].c_{i}\neq\CStrue\} } {} \label{eq:Aa_ask} \\
        %now
        & \Aa{\anow{d}{A}{B}}{\I[]} = \{\crs{(d,\emptyset)}{d}{d}\cdot\ed \mid
        \ed \in \Aa{A}{\I[]}\} \sqcup \notag\\*&\qquad \lubC{}{}\{\crs{(c^+\CSmerge
        d,c^{-})}{c\CSmerge d}{c' \CSmerge d}\cdot (\prop{d}{s}) \mid
        \begin{aligned}[t]
            &\crs{(c^+,c^{-})}{c}{c'}\cdot s \in \Aa{A}{\I[]}, \\
            &c\CSmerge d\soddcond (c^{+}\CSmerge d,c^{-})\}\sqcup
        \end{aligned}
        \notag\\*&\qquad \lubC{}{}\{\crs{(d,C)}{d}{d} \cdot (\prop{d}{s}) \mid
        \stutt{C}\cdot s \in \Aa{A}{\I[]},\, d\soddcond (d,C)\}\sqcup
        \notag\\*&\qquad \lubC{}{}\{\crs{(\CStrue,d)}{\CStrue}{\CStrue}\cdot\ed \mid \ed
        \in \Aa{B}{\I[]} \} \sqcup
        \notag\\*&\qquad \lubC{}{}\{\crs{(c^+,c^{-}\cup \{d\})}{c}{c'}\cdot s \mid
        \begin{aligned}[t]
            & \crs{(c^+,c^{-})}{c}{c'}\cdot s \in \Aa{B}{\I[]}, \\
            & c\soddcond (c^{+},c^{-}\cup \{d\}) \}\sqcup
        \end{aligned}
        \notag\\*&\qquad \lubC{}{}\{\crs{(\CStrue, C\cup
        \{d\})}{\CStrue}{\CStrue}\cdot s \mid \stutt{C}\cdot s \in \Aa{B}{\I[]} \}
        \label{eq:Aa_now} \\
        %parallel
        & \Aa{A \parallel B}{\I[]}=\lubC{}{}\big\{\parallelseq{s_{A}}{s_{B}} \mid
        s_{A}\in \Aa{A}{\I[]}, s_{B}\in \Aa{B}{\I[]} \big\}
        \label{eq:Aa_parallel} \\
        %exists
        & \Aa{\ahiding{x}{A}}{\I[]}=\lubC{}{}\big\{ s \in \C \mid
        \begin{aligned}[t]
            & \exists s' \in \Aa{A}{\I[]} \text{ such that }
            \exists_{x}s=\exists_{x}s',\\
            & s' \text{ is $x$-connected, $s$ is $x$-invariant}\}
        \end{aligned}
        \label{eq:Aa_hiding} \\
        %procedure call
        & \Aa{ p(z)}{\I[]}=\lubC{}{}\big\{\crs{(\CStrue,\emptyset)}{\CStrue}{\CStrue}\cdot s\
        \mid s\in \I[](p(z))\big\}
        \notag %\label{eq:Aa_pcall}
    \end{align}
}%
\end{definition}

Let us now illustrate the idea of the semantics.
The \atell{} agent works independently of the current store, thus in
\eqref{eq:Aa_tell}, the conditional reactive sequence starts with a
conditional tuple composed by the condition $(\CStrue,\emptyset)$, which is
always satisfied, and a
second part that says that the constraint $c$ is added during the first
computational step;
afterwards, the computation terminates with $\ed$.  

The semantics for the non-deterministic choice \eqref{eq:Aa_ask}, collects
for each guard $c_i$ a conditional sequence of the form
$\crs{(c_{i},\emptyset)}{c_{i}}{c_{i}}\cdot (\prop{c_{i}}{s})$.  The
condition states that $c_i$ has to be satisfied by the current store,
whereas the pair $\langle c_i, c_i \rangle$ represents the fact that the
query to the store does not modify the store.  The constraint $c_i$ is
propagated
into the sequence $s$ (the continuation of the computation which belongs to
the semantics of $A_i$) by means of the propagation operator that
(consistently) adds a given constraint to the stores appearing in a
sequence: \par\vskip -2ex {\small
\begin{equation*}
 	\prop{h}{s} =
 		\begin{cases}
			\crs{\eta}{a\CSmerge h}{b \CSmerge h}\cdot(\prop{h}{s'})
			& \text{if}\ s=\crs{\eta}{a}{b}\cdot s',\ \eta^+\CSmerge h\neq\CSfalse,\\
			& b\CSmerge h\neq\CSfalse,%\
a\CSmerge h\soddcond \eta \\[0.5ex]
			\crs{\eta}{a\CSmerge h}{\CSfalse}
			& \text{if}\ s=\crs{\eta}{a}{b}\cdot s',\ \eta^+\CSmerge h\neq\CSfalse,\\
			& b\CSmerge h=\CSfalse,%\ \text{and}\ 
a\CSmerge h
\soddcond \eta\\[0.5ex]
			\stutt{\eta^{-}}\cdot(\prop{h}{s'}) &  \text{if }s=\stutt{\eta^{-}}\cdot s'\\[0.5ex]
			s & \text{if}\ s=\ecrs\ \text{or}\ s=\ed
 		\end{cases}
 	\end{equation*}
}%
In addition, we have
to model the case when the computation suspends, \ie{} when no guard of the
agent is satisfied by the current store.  Sequences representing this
situation are of the form $\stutt{\cup_{i=1}^{n} \{c_{i}\} } \cdot s$ where
$s$ is, recursively, an element of the semantics of the choice agent.
The only case when we do not include the stuttering sequence is when one of
the guards $c_i$ is $\CStrue$.  Note that, due to the partial nature of the
constraint system, the fact that the disjunction of the guards is $\CStrue$
is not a sufficient condition to avoid %the 
suspension.

The definition of the conditional agent $\anow{}{}{}$ is similar to the
previous one.  However, since it is instantaneous, we have 6 cases
depending on the 3 possible heads of the sequences of the semantics of $A$
(\resp\ $B$) and on the fact that the guard $d$ is satisfied or not in the
current time instant.

The semantics for the parallel composition of two agents
\eqref{eq:Aa_parallel}, is defined in terms of an auxiliary commutative
operator $\parallelseq{}{}$ which combines the sequences of the two agents:
\par\vskip -3ex {\small
\begin{equation*}
	\parallelseq{s_{A}}{s_{B}}=
		\begin{cases}
		\begin{aligned}[t]
		\crs{(\eta\uncond \delta)&}{a\CSmerge c}{b\CSmerge d}\cdot%\\
\parallelseq{(\prop{d}{s_{A}'})}{(\prop{b}{s_{B}'})}
		\end{aligned}\!\!
		&\text{if}\
		\begin{aligned}[t]
		&s_{A}=\crs{\eta}{a}{b}\cdot s_{A}',\\
		& s_{B}=\crs{\delta}{c}{d}\cdot s_{B}',\\
		&a\CSmerge c \soddcond (\eta\uncond \delta), %\ \text{and}\ 
b\CSmerge d \neq\CSfalse
		\end{aligned}\\[0.5ex]
		\crs{(\eta\uncond \delta)}{a\CSmerge c}{\CSfalse}
		&\text{if}\
		\begin{aligned}[t]
		&s_{A}=\crs{\eta}{a}{b}\cdot s_{A}',\\
		& s_{B}=\crs{\delta}{c}{d}\cdot s_{B}',\\
		&a\CSmerge c \soddcond (\eta\uncond \delta), %\ \text{and}\ 
b\CSmerge d =\CSfalse   
		\end{aligned}\\[0.5ex]   
		\begin{aligned}[t]
		\crs{(\eta^{+},\eta^{-}\cup\delta^{-})&}{a}{b}\cdot%\\
		%&
\parallelseq{s_{A}'}{(\prop{b}{s_{B}'})}
		\end{aligned}
		&\text{if }\
		\begin{aligned}[t]
		&s_{A}=\crs{\eta}{a}{b}\cdot s_{A}',\\
		&s_{B}=\stutt{\delta^{-}}\cdot s_{B}',\\
		&%\text{and}\ 
a\soddcond (\eta^{+},\eta^{-}\cup\delta^{-})
        \end{aligned}\\[0.5ex]       
		\stutt{\eta^{-}\cup \delta^{-}} \cdot
		\parallelseq{s_{A}'}{s_{B}'}
		&\text{if}\
		\begin{aligned}[t]
		&s_{A}'=\stutt{\eta^{-}}\cdot s_{A}',\\
		&%\text{and}\ 
s_{B}'=\stutt{\delta^{-}}\cdot s_{B}'
		\end{aligned}\\[0.5ex]
	s_{A} & \text{if }s_{B}= \ecrs \text{ or }s_{B}=\ed
	\end{cases}
\end{equation*}       
}%

For the hiding operator \eqref{eq:Aa_hiding}, we collect the
sequences that satisfy the restrictions regarding the visibility of the
hided variables.
In particular, a conditional reactive sequence $s=t_1\dots t_n\dots$
    is $x$\emph{-connected} when
(1) if $t_1=\crs{\eta_1}{a_1}{b_1}$ then $\exists_{x}a_{1}=a_{1}$ 
and (2)
for each $t_i=\crs{\eta_i}{a_i}{b_i}$ and
$t_{i+1}=\crs{\eta_{i+1}}{a_{i+1}}{b_{i+1}}$, with $i>1$,
$\exists_{x}a_{i+1}\CSmerge b_{i}=a_{i+1}$.
    A conditional reactive sequence $s=t_1\dots t_n\dots$ is
$x$\emph{-invariant} if for each
    computational step $t_i=\eta_i \ra \langle a_i,b_i \rangle$, it holds that
$b_i=\exists_{x}b_i\CSmerge a_i$.

Finally, the semantics of the process call $\apcall{p}{\vec{x}}$
collects the sequences in the interpretation $\I[](\apcall{p}{\vec{x}})$,
delayed by one time unit, as stated in the operational semantics.

Let us show an illustrative example. Consider the \tccp{} agent
$A\equiv\aask{y\geq0} \atell{z\leq0}$. The semantics is composed of two %one
sequences: 
\par\vskip -3ex {\small
\begin{align*}
    \Aa{A}{\I[]}=&\{
    \begin{aligned}[t]		
        & \crs{(y\geq0,\emptyset)}{y\geq0}{y\geq0}\cdot
        \crs{(\CStrue,\emptyset)}{y\geq0}{ y\geq0\CSmerge z\leq0}\cdot \ed\}
        \end{aligned}\\ %[1ex]
        &\cup\{\stutt{y\geq0}\cdot s \mid s
        \in\Aa{A}{\I[]}\}%\aask{y\geq0}\atell{z\leq0}
\end{align*}
}%

\subsection{Fixpoint Denotations of Declarations}\label{sec:Fss}

\begin{figure}
    \scriptsize
    \begin{center}
        \begin{tikzpicture}[scale=0.55]
            \draw (0,0) coordinate (root); 
            \draw (root)+(0,-1.5) node (one)
            {$\begin{aligned}&(\exists_{y}(y>x),\emptyset)\ra\\&\pair{\exists_{y}(y>x)}{
            \exists_{y}(y>x)}\end{aligned}$};
            \draw (root)+(-7,-1) node (two)
            {$\begin{aligned}&(\exists_{y}(y\leq x),\emptyset)\ra\\&\pair{\exists_{y}(y\leq
            x)}{\exists_{y}(y\leq x)}\end{aligned}$};
            \draw (root)+(7,-1) node (stutt) {$\stutt{\{y>x\}\cup\{y\leq
            x\}}$};   
            \draw (two)+(0,-2) node (three) {$\ed$};
            \draw (one)+(-2,-2) node (four)
            {$\begin{aligned}&(\CStrue,\emptyset)\ra\\&\pair{\exists_{y}(y>x)}{\exists_{y}
            (y>x)}\end{aligned}$};
            \draw (four)+(0,-2) node (seven)
            {$\begin{aligned}&(\exists_y(y>x+1),\emptyset)\ra\\&\pair{\exists_{y}(y>x+1)}{
            \exists_{y}(y>x+1)}\end{aligned}$};
            \draw (seven)+(-5,-2.5) node (eight)
            {$\begin{aligned}&(\CStrue,\emptyset)\ra\\&\pair{\exists_{y}(y>x+1)}{\exists_{y}(y>x+1)}\end{aligned}$};
            \draw (eight)+(0,-1.5) coordinate (fourteen); 
            \draw (seven)+(5,-2.5) node (nine)
            {$\begin{aligned}&(\exists_{y}(y\leq
            x+2),\emptyset)\ra\\&\pair{\exists_{y}(y=x+2)}{\exists_{y}(y=x+2)}\end{aligned}$
            };
            \draw (one)+(5,-3) node (five)
            {$\begin{aligned}&(\exists_{y}(y\leq
            x+1),\emptyset)\ra\\&\pair{\exists_{y}(y=x+1)}{\exists_{y}(y=x+1)}\end{aligned}$
            };
            \draw (five)+(0,-2) node (six) {$\ed$};
            \draw (nine)+(0,-1.5) node (ten) {$\ed$};
            \draw (stutt.south)+(1.5,-1.5) coordinate (t1);
            \draw (stutt.south)+(-1.5,-1.5) coordinate (t2);
            \draw (stutt.south) -- (t1) -- (t2) -- cycle;       
            \draw (stutt.south)+(0,-1) node (Aask) {$\Aa{\aask{}}{\I[]}$};   
            
            %% information pre-order
            \draw[-] (root) -- (stutt); 
            \draw[-] (root) -- (one);    
            \draw[-] (root) -- (two); 
            \draw[-] (two) -- (three); 
            \draw[-] (one) -- (four); 
            \draw[-] (one) -- (five); 
            \draw[-] (five) -- (six);
            \draw[-] (four) -- (seven); 
            \draw[-] (seven) -- (eight); 
            \draw[-] (seven) -- (nine); 
            \draw[-] (nine) -- (ten); 
            \draw[densely dashed] (eight) -- (fourteen); 
        \end{tikzpicture}
    \end{center}
\vspace{-4ex}
    \caption{Tree representation of $\F[]{D}$ in the example.}
    \label{fig:tree2}
\end{figure}
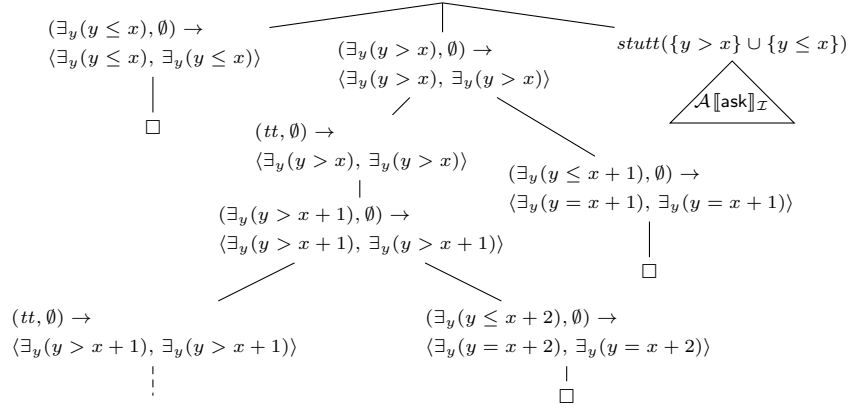

Now we can define the semantics for a set of process declarations $D$ as
the fixpoint of the immediate consequences operator
    $\Dd{D}{\I[]} \dfn \lambda p(x).  \lubC{_{p(x) \clauseif A\in D}
    \Aa{A}{\I[]}} {}$, which is continuous.
Thus, it has a least fixpoint and we can define the semantics of $D$ as
	$\F[]{D}=\lfp(\Dd{D}{})$.
As an example, in Figure~\ref{fig:tree2} we represent the (infinite) set of
traces of $\F[]{\{p(x) \clauseif \ahiding{y}( \aask{y>x} p(x+1) +
\aask{y\leq x}\askip)\}}$.%
\footnote{For the sake of simplicity, we assume that we can use expressions
of the form $x+1$ directly in the arguments of a process call.  We can
simulate this behavior by writing $\atell{x'=x+1}\ra p(x')$ (but
introducing a delay of one time unit).  }

In \cite{CominiTV11sem} we have proven that $\BehaEq{D_1}{D_2}$ if and only if
$\F[]{D_1} =\F[]{D_2}$ (correctness and full abstraction of $\F[]{}$ \wrt\
$\BehaEq{}{}$).

\section{Abstract semantics for \tccp: the abstraction scheme} 
\label{sec:abstr}

In this section, starting from the fixpoint semantics in Section~\ref{sec:Sem},
we present an abstract semantics which approximates the observable behavior of
the program. Program properties that are of interest are Galois Insertions
between the concrete domain and the chosen abstract domain. We assume
familiarity with basic results of abstract interpretation \cite{CousotC79}.

We define an abstraction scheme where we develop the abstraction of
computations, \ie\ of maximal sets of conditional reactive sequences, by
successive lifting.  We start with a function that abstracts the
information component of the program semantics, \ie\ the store; then we
build the abstraction of conditional tuples; then of conditional reactive
sequences and, finally, of maximal sets.

We start from an upper-approximating function $\alpiu{} : \CSdom \to
\ACSdom$ into an abstract constraint system $\ACSys = \langle \ACSdom,
\ACSord, \ACSmerge, \ACSjoin, \ACStrue, \ACSfalse, \Var, \ACShid \rangle$,
where $\ACStrue$ and $\ACSfalse$ are the smallest and the greatest abstract
constraint, \resp.  We often use the inverse relation $\ACSimp$ of
$\ACSord$.
We have also a lower-approximating function $\almeno{} : \wp(\CSdom) \to
\ACSdomDual$ into an abstract constraint system $\ACSysDual = \langle
\ACSdomDual, \ACSordDual, \ACSmergeDual, \ACSjoinDual, \ACStrueDual,
\ACSfalseDual, \Var, \ACShidDual{} \rangle$.  This second function is needed to
(correctly)
deal with the negative part of conditions.

We have two ``external'' operations $\ACSinj{}{} : \CSdom \times \ACSdom
\to \ACSdom$ and $\ACSinjDual{}{} : \CSdom \times \ACSdomDual \to
\ACSdomDual$ that update an abstract store with a concrete constraint
(coming from the program).  In addition, a ``bridge'' relation
$\ACSimppn\ \in\ \ACSdom \times \ACSdomDual$ decides if an upper-abstract
constraint is consistent with a lower-abstract constraint.
Abstract and concrete constraint systems are related by these conditions:
\par\vskip -3ex {\small
\begin{align*}
  & \ACSinj {c} {\alpiu{a}} = \alpiu{c \CSmerge a}
        %\\ 
        &  \ACSinjDual {c} {\almeno{C}} = \almeno{ \{c\} \cup C} 
        \\
        & \alpiu{a \CSmerge b} = \alpiu{a} \ACSmerge \alpiu{b}
        %\\    
        &  \almeno{C \cup C'} = \almeno{C} \ACSjoinDual \almeno{C'} 
        \\
        & a \CSimp b \Longrightarrow  \alpiu{a} \ACSimp \alpiu{b} 
	   %\\
        & \almeno{\{a\}} \ACSimpDual \almeno{C} \Longrightarrow \exists c \in
        C .\, a \CSimp c
        \\
        & \alpiu{\CShid[x]{a}} = \ACShid[x]{ \alpiu{a}}
        %\\
        & \almeno{\{ \CShid[x]{c} \mid c\in C\}} = \ACShidDual[x]{\almeno{C}}
        \\
        & \forall c\in C.\ a\not\CSimp c \Longrightarrow
        \alpiu{a}\not\ACSimppn\almeno{C}
\end{align*}
}%

An \emph{abstract condition} is a pair %of abstract constraints 
of the form $(\ACSetap,\ACSetam) \in \ACSdom \times \ACSdomDual$.
Similarly to the concrete case, given an abstract condition $\ACSetac =
\ACSetapm$ and an abstract store $\ACSc \in \ACSdom$, we say that $\ACSc$
satisfies $\ACSetac$ (written $\ACSc \Asoddcond \ACSetac$) when
$\ACSetap\neq\ACSfalse$ and $\ACSc\ACSimp\ACSetap$, but $\ACSc\not\ACSimppn
\ACSetam$.
Given an abstract condition $\ACSetac$,
$\ACSc[a],\ACSc[b] \in \ACSdom$ and $\ACScDual \in \ACSdomDual$, %we define
an abstract conditional tuple is either a triple
$\Acrs[m]{\ACSetac}{\ACSc[a]}{\ACSc[b]}$, such that $\ACSc[a]\Asoddcond
\ACSetac$, or a construct of the form $\Astutt[m]{\ACScDual}$, where
$m\in\{0,+\infty\}$ states how many times the corresponding %concrete
tuples appear consecutively in the %concrete 
sequence.
Given a (concrete) conditional tuple $t$, we define its abstraction $\al{t}$ as
\par\vskip -3ex {\small
\begin{align*}	   
    &\al{\crs{(\eta^{+},\eta^{-})}{a}{b}} =
    \AcrsC[1]{\alpiu{\eta^{+}}}{\almeno{\eta^{-}}}{\alpiu{a}}{\alpiu{b}}
    \\
    &\al{\stutt{C}} = \Astutt[1]{\almeno{C}}
\end{align*}
}%

Now, an abstract conditional reactive sequence is a sequence of
\emph{different} abstract tuples $\tilde{t}_1\dots\tilde{t}_m\dots$, maybe
ended with $\ed$.
The natural number associated to each abstract conditional tuple is needed
to keep synchronization among processes due to the particularly strong synchronization
properties of the language, as already noticed in \cite{AlpuenteGPV05}.

The abstraction $\al{s}$ of a sequence of conditional tuples $s$
is defined by structural induction on the form of its tuples. % $t$.
It collapses all the computation steps (conditional tuples) that, after
abstraction, coincide.  Formally, $\al{\ecrs}=\ecrs$, $\al{\ed}=\ed$ and
\begin{align*}
    & \al{t\cdot r} \dfn
    \begin{cases}
        \Acrs[m+1]{\ACSetac}{\ACSc[a]}{\ACSc[b]} \cdot \tilde{r} & \text{if
        $\al{t} = \Acrs[1]{\ACSetac}{\ACSc[a]}{\ACSc[b]}$, $\al{r} =
        \Acrs[m]{\ACSetac}{\ACSc[a]} {\ACSc[b]} \cdot \tilde{r}$} \\
        \Astutt[m+1]{\ACScDual} \cdot \tilde{r} & \text{if $\al{t} =
        \Astutt[1]{\ACScDual}$, $\al{r} = \Astutt[m]{\ACScDual} \cdot
        \tilde{r}$} \\
        \al{t}\cdot \al{r} & \text{otherwise}
    \end{cases}
\end{align*}
We extend this definition to sets of conditional sequences in the natural
way.

We denote by $\A$ the domain $\al{\M}$ of the sets of abstract
conditional reactive sequences.  By adjunction we derive the concretization
function $\ga{}$ such that $$\latticeC \galoiS{\al{}}{\ga{}} \latticeA$$%,
where $a \leqA a' \Longleftrightarrow \ga{a} \leqC \ga{a'}$.

This abstraction can be systematically lifted to the domain of
interpretations: $\Ii\galoiS{\al{}}{\ga{}} [\MGC \ra\A]$ so that we can
derive the optimal abstraction of $\Dd{D}{}$ simply as
$\ADd{D}{}:=\al{}\circ\Dd{D}{}\circ\ga{}$.  The abstract interpretation
theory ensures that $\F{D}:=\ADd{D}{}\uparrow\omega$ is the best correct
approximation of $\F[]{D}$.

It turns out that $\ADd{D}{\I} = \lambda p(x).\lubA{_{p(x) \clauseif A\in
D} \AAa{A}{\I}} {}$, where $\AAa{\cdot}{\I}$ is defined by structural
induction on the syntax in a similar way as the concrete version.
Given the similarity to the concrete case, in the following we describe
only two cases in order to illustrate the use of the upper- and
lower-approximations 
(for full details consult~\cite{CominiTV11sem}). 
The semantics for the $\atell{}$ agent just applies the abstraction to the
only
concrete sequence, thus: $\AAa{\atell{c}}{\I} =
\{\Acrs[1]{(\ACStrue,\ACSfalseDual)}{\ACStrue} {\alpiu{c}} \cdot\ed\}$.
For the $\anow{}{}{}$ semantics, % must consider several cases, similarly to
we only show the general case when the condition holds, and the general
case when it does not hold:
\par\vskip -3ex {\small
\begin{align*}
    & \AAa{\anow{d}{A}{B}}{\I} = \\*&\quad
    \{\AcrsC[n]{\ACSinj{d}{\ACSetap}}{\ACSetam}{\ACSinj{d}{\ACSc[a]}}{\ACSinj{d}{\ACSc[b]}}
    \!\cdot\! (\Aprop{d}{\tilde{s}}) \mid % {} %\\*&\qquad \qquad
    \AcrsC[n]{\ACSetap}{\ACSetam}{\ACSc[a]}{\ACSc[b]}\!\cdot\! \tilde{s} \!\in\!
    \AAa{A}{\I}, %\text{ and } 
    \ACSinj{d\!}{\!\ACSc} \!\Asoddcond\!
    (\ACSinj{d\!}{\!\ACSetap},\ACSetam)\} \\*&\quad
    \lubA{}{\ldots}\lubA{}{~} \\*&\quad
    \{\AcrsC[1]{\ACSetap}{\ACSinjDual{d}{\ACSetam}}{\ACSc[a]}{\ACSc[b]}
    \!\cdot\! \AcrsC[n]{\ACSetap}{\ACSetam}{\ACSc[a]}{\ACSc[b]}\!\cdot\! \tilde{s}
    \!\mid\! %{} \\*&\qquad \qquad
    \AcrsC[n+1]{\ACSetap}{\ACSetam}{\ACSc[a]}{\ACSc[b]}\!\cdot\! \tilde{s}
    \!\in\!\AAa{B}{\I}, %\text{ and } 
    \ACSc\!\Asoddcond\! (\ACSetap,\ACSinjDual{d\!}{\!\ACSetam}) \} \\*&\quad \lubA{}{\ldots}
\end{align*}
}
the $\Aprop{}{}$ operator is the abstract counterpart of the concrete
version.

\section{Abstract diagnosis of timed concurrent constraint programs}
\label{sec:abs-diag}

Now, following the ideas of \cite{CominiLMV96a}, we define the abstract
diagnosis of \tccp. %timed concurrent constraint programs.  
The framework of abstract diagnosis \cite{CominiLMV96a} comes from the idea
of considering the abstract versions of Park's Induction
Principle\footnote{A concept of formal verification that is undecidable in
general.}.  It can be considered as an extension of declarative debugging
since there are instances of the framework that deliver the same results.
In the general case, diagnosing \wrt\ \emph{abstract} %program UNDERFULL
properties relieves the user from having to specify in excessive detail the
program behavior (which could be more error-prone than the coding itself).

Let us now introduce the workset of abstract diagnosis.  Having chosen a
property of the computation $\alpha$ of interest (an instance of the
abstraction scheme of Section~\ref{sec:abstr}), given a set of declarations
$\cR$ and $\Sz \in \A$, which is the specification of the intended behavior
of $D$ \wrt\ the property $\alpha$, we say that
    \begin{enumerate}
        
        \item\label{pt:COR:def:Correct}
        $\cR$ is (abstractly) \emph{partially correct}\index{partially
        correct} \wrt\ $\Sz$ if $\alpha(\F[]{\cR}) \leq \Sz$.
        
        \item\label{pt:COM:def:Correct}
        $\cR$ is (abstractly) \emph{complete}\index{complete} \wrt\ $\Sz$
        if $\Sz \leq \alpha(\F[]{\cR})$.
        
        \item
        $\cR$ is \emph{totally correct}\index{totally correct} \wrt\ $\Sz$,
        if it is partially correct and complete.
    \end{enumerate}
In this setting, the user can only reason in terms of the properties of the
expected concrete semantics without being concerned with (approximate)
abstract computations.
The \emph{diagnosis} determines the ``originating'' symptoms and, in the
case of incorrectness, the relevant process declaration in the program.
This is captured by the definitions of \emph{abstractly incorrect process
declaration} and \emph{abstract uncovered element}:
\begin{definition}
    \label{def:ab.incorclau}
    \label{def:ab.uncovered}
    
    Let $\cR$ be a set of declarations, $R$ a process declaration and $\{e\},
    \Sz \in \A$.
    
    $R$ is \emph{abstractly incorrect} \wrt\ $\Sz$ if $\Tp{\{R\}}{\Sz}
    \not\leq \Sz$.
    
    $e$ is an \emph{uncovered element} \wrt\ $\Sz$ if $\{e\} \leq \Sz$ and
    $\glbA{\{e\}}{\Tp{\cR}{\Sz}} = \botA$.
\end{definition}
Informally, $R$ is abstractly incorrect if it derives a wrong abstract
element from the intended semantics.
$e$ is uncovered if the process declarations cannot derive it from the
intended semantics.

It is worth noting that the notions of 
correctness and completeness are defined
in terms
of $\alpha(\F[]{\cR})$, \ie{} in terms of abstraction of the concrete
semantics.  The abstract version of algorithmic debugging \cite{Shapiro82},
which is based on symptoms (\ie\ deviations between $\alpha(\F[]{\cR})$ and
$\Sz$), requires the construction of $\alpha(\F[]{\cR})$ and therefore a
fixpoint computation.
In contrast, the notions of abstractly incorrect process declarations and
abstract
uncovered elements are defined in terms of \emph{just one} application of
$\Tp{\cR}{}$ to $\Sz$.
The issue of the precision of the abstract semantics is specially
relevant in establishing the relation between the two concepts
(\ie{} %of proving which is 
the relation between abstractly incorrect
process declarations and abstract uncovered elements on one side, and
abstract partial correctness and completeness, on the other
side).\footnote{Proofs are available at
\texttt{http://www.dimi.uniud.it/comini/Papers}.}

\begin{theorem}\label{th:ab.corr-compl}
    \begin{enumerate}
        
        \item\label{pt:ab.correct} If there are no abstractly incorrect
        process declarations in $\cR$, then $\cR$ is partially correct
        \wrt\ $\Sz$.
        
        \item\label{pt:ab.complete2} Let $\cR$ be partially correct \wrt\
        $\Sz$.  If $\cR$ has abstract uncovered elements then $\cR$ is not
        complete.
    \end{enumerate}
\end{theorem}

When applying the diagnosis \wrt\ approximate properties, the results  may be
weaker than those that can be achieved on concrete domains just because of
approximation.
Abstract incorrect process declarations are in general just a warning about
a possible source of errors.  Because of the approximation, it can happen
that a (concretely) correct declaration is abstractly incorrect.
However, as shown by the following theorem, all concrete errors 
are detected, as they lead to an abstract incorrectness or abstract
uncovered.%
\begin{theorem}
    \label{th:abs-conc}
    
    Let $r$ be a process declaration and $\Sz[]$ a concrete specification.
    \begin{enumerate}
        
        \item\label{pt:new} If %{\small
$\Tp[]{\{r\}}{\Sz[]} \!\!\not\leqC\! \Sz[]$
%} 
and
        %{\small 
$\alpha(\Tp[]{\{r\}}{\Sz[]}) \!\!\not\leq\! \alpha(\Sz[])$
%}
then $r$ is
        abstractly incorrect \wrt\ {\small $\alpha(\Sz[])$}.
    
        \item\label{pt:newunc} If there exists an abstract uncovered
        element $a$ \wrt\ $\alpha(\Sz[])$, such that $\gamma(a) \leqC
        \Sz[]$ and $\gamma(\botA) = \botA$, then there exists a concrete
        uncovered element $e$ \wrt\ $\Sz[]$ (\ie\ $e \leqC \Sz[]$ and $e
        \sqcap \Tp[]{\cR}{\Sz[]} = \botC$).
    \end{enumerate}
\end{theorem}

It is particularly
useful for applications the fact that our proposal can be used with partial
specifications and also with partial programs.  Obviously, one cannot
detect errors in process declarations involving processes which have not
been specified, but for the process declarations that involve %only
processes that have a specification, the check can be made, even if the
whole program has not been written yet.  This includes the possibility of
applying our ``local'' method to all parts of a program not involving
constructs which we cannot handle (yet).  With other ``global'' approaches
such programs could not be checked at all.

It is worthy to note that, even for a noetherian abstract constraint system
$\ACSys$, the domain of abstract sequences defined above is not---in
general---noetherian, due to the use of the index in each tuple (we cannot
get rid of it since it is needed to keep synchronization among parallel
processes).
This means that our current proposal cannot be used for static program
analysis, unless we resort to use widening operators.  However (for
noetherian abstract constraint systems) our abstract diagnosis is effective
since specifications have to be abstractions of some concrete semantics
and, since the store evolves monotonically, it holds that the number of
conditional tuples that can appear in an abstract sequence is, thus, finite.

\subsection{Examples of application of the framework}

Let us now show two illustrative examples of the approach. 
The first example shows the new ability of our approach:
that of dealing with the constructors that introduce the
non-monotonic
behavior
of the system, in particular the $\anow{}{}{}$ agent. 

\begin{example}\label{ex:negative-info}
    We model a (simplified) \emph{time-out}$(n)$ process
    that checks for, at most, $n$ times units if the system emits a signal
    telling that the process evolves normally ($\mathit{system = ok}$).
    When the signal arrives, the system emits the fact that there is no alert
    ($\mathit{alert}=\mathit{no}$)\footnote{The classical timeout would restart the
    countdown by recursively calling \emph{time-out}$(n)$.}.
    Let $d_0$, $d_n$, $d_{\mathit{action}}$ be the following 
    declarations:
\par\vskip -3ex {\small
\begin{align*}
           \text{\em time-out}(0) \clauseif
        &\anow{\mathit{system=ok}}{\mathit{action}}{}\ \mathsf{else}%\\
        %&
\ (\aask{\CStrue}{\text{\em time-out}(0)})\\%[1ex] 
        \text{\em time-out}(n) \clauseif
        &\anow{\mathit{system=ok}}{\mathit{action}}{}\ \mathsf{else}%\\
\ (\aask{\CStrue} \text{\em time-out}(n-1))\\%[1ex]
        \mathit{action}\clauseif & \atell{\mathit{alert}=\mathit{no}}%
    \end{align*}
}%
When the time limit is reached (declaration $d_0$), the system should set the
signal
    $\mathit{alert}$ to \emph{yes} ($\atell{\mathit{alert}=\mathit{no}}$). 
However, we have introduced an error in the 
program, calling the process recursively instead: $\text{\em time-out}(0)$.

    Due to the simplicity of the constraint system, the abstract
domain coincide with the concrete one, and the two external 
functions are the $\ACSjoin$ and $\ACSjoinDual$ operators.
    
Let us now consider the following specification.
For $d_0$ we expect that, if the $\mathit{ok}$ signal is present, 
then it ends with an $\mathit{alert=no}$ signal, otherwise an alert should be emitted. 
This is represented by two possible sequences, one with a condition where $\mathit{system=ok}$,
and a second one when $\mathit{system=ok}$ is absent (this is a sequence that
reasons with the \emph{absence} of information).
\par\vskip -3ex {\small    
\begin{align*}
         &\Sz(\text{\em time-out}(0)) = \begin{aligned}[t]
                                         &\{\begin{aligned}[t]
                                            &\Acrs[1]{(\mathit{system=ok},\ACSfalseDual)}{\mathit{system=ok}}{\mathit{system=ok}}
                                             \cdot\\
                                            &\Acrs[1]{(\ACStrue,\ACSfalseDual)}{\mathit{system=ok}}{\mathit{system=ok}\ACSmerge
                                             \mathit{alert=no}}\cdot \ed\}
                                         \end{aligned}\\
                                        &{}\cup %\begin{aligned}[t]
                                                \{\Acrs[1]{(\ACStrue,\{\mathit{system=ok}\})}{\ACStrue}{\ACStrue}\cdot
                                                \Acrs[1]{(\ACStrue,\ACSfalseDual)}{\ACStrue}{\mathit{alert=yes}}\cdot \ed\}
                                        \end{aligned}%\\%[1ex]
\end{align*}
}%
The specification for $d_n$ is similar, but we add $n$ sequences, since we have the possibility that 
the signal arrives at each time instant before $n$.
\par\vskip -3ex {\small    
\begin{align*}
         &\Sz(\text{\em time-out}(n)) = \begin{aligned}[t]
                                        &\{\begin{aligned}[t]
                                           &\Acrs[m]{(\ACStrue,\{\mathit{system=ok}\})}{\ACStrue}{\ACStrue}\cdot\\
                                           &\Acrs[1]{(\mathit{system=ok},\ACSfalseDual)}{\mathit{system=ok}}{\mathit{system=ok}}
                                           \cdot\\
                                           &\Acrs[1]{(\ACStrue,\ACSfalseDual)}{\mathit{system=ok}}{\mathit{system=ok}\ACSmerge
                                           \mathit{alert=no}}\cdot \ed \mid 0\leq m<n\}
                                           \end{aligned}\\
                                        &{}\cup\begin{aligned}[t]           
                                               \{&\Acrs[n+1]{(\ACStrue,\{\mathit{system=ok}\})}{\ACStrue}{\ACStrue}\cdot\Acrs[1]{(\ACStrue,\ACSfalseDual)}{\ACStrue
                                                  }{\mathit{alert=yes}}\cdot \ed\}
                                                \end{aligned}\\%[1ex]
                                        \end{aligned}\\
         &\Sz(\mathit{action})=\{\Acrs[1]{(\ACStrue,\ACSfalseDual)}{\ACStrue}{\mathit{alert=no}}\cdot
         \ed\}
    \end{align*}
}	
    Now, when we compute $\Tp{\{d_0\}}{\Sz}$ we have:
\par\vskip -3ex {\small
    \begin{align*}
        &
        \begin{aligned}[t]
            \{&\Acrs[1]{(\mathit{system=ok},\ACSfalseDual)}{\mathit{system=ok}}{\mathit{system=ok}}
            \cdot\\
            &
            \Acrs[1]{(\ACStrue,\ACSfalseDual)}{\mathit{system=ok}}{\mathit{system=ok}\ACSmerge
            \mathit{alert=no}}\cdot \ed\}
        \end{aligned}\\
        &{}\cup
        \begin{aligned}[t]
            \{&\Acrs[1]{(\ACStrue,\{\mathit{system=ok}\})}{\ACStrue}{\ACStrue}\cdot
\Acrs[1]{(\mathit{system=ok},\ACSfalseDual)}{\mathit{system=ok}}{\mathit{system=ok}}
\cdot\\
            &\Acrs[1]{(\ACStrue,\ACSfalseDual)}{\mathit{system=ok}}{\mathit{system=ok}\ACSmerge
\mathit{alert=no}}\cdot \ed\}
        \end{aligned}\\
        &{}\cup
        \begin{aligned}[t]
            \{&\Acrs[2]{(\ACStrue,\{\mathit{system=ok}\})}{\ACStrue}{\ACStrue}\cdot
\Acrs[1]{(\ACStrue,\ACSfalseDual)}{\ACStrue}{\mathit{alert=no}}\cdot \ed\}
        \end{aligned}
    \end{align*}
}%
    Due to the last sequence, $\Tp{\{d_0\}}{\Sz}{\not\leqA} \Sz$, so we
    conclude that $d_0$ is (abstractly) incorrect. This is due to the recursive
    call in the else branch of the declaration. 
If we fix the program replacing $d_0$ by $d'_{0}$ where
    the recursive call is replaced by $\atell{\mathit{alert}=\mathit{yes}}$, then
    $\Tp{\{d'_0\}}{\Sz} \leq \Sz$, thus $d'_{0}$ is abstractly correct.
\end{example}

In \cite{FalaschiOPV07} it was studied an example where a \emph{control} process
checks 
whether a \emph{failure} signal arrives to the system. The most important point
that differs from the \emph{timeout} example is that, in the \emph{control}
case,
someone has to explicitly tell the system that an error has occurred.
Instead, in the \emph{timeout} example, the system is able to act (and maybe
recover)
when it detects that something that should have happened, hadn't. 
In other words, the \emph{control} example does not
handle \emph{absence} of information, since non-monotonic operators are not
considered there. We have implemented the example in \tccp{} %by using streams
and we have checked that the same results can be achieved in our framework
if we apply the same abstraction they use (a \depthk{} abstraction).

The second example %Example~\ref{ex:abstractionCS} shows 
we show illustrates how one can work with the abstraction
of the constraint system, and also how we can take advantage of our
abstract domain.

\begin{example}\label{ex:abstractionCS}
    Let us consider a system with a single declaration and the abstraction
    of the constraint system that abstracts integer variables to a
    (simplified) interval-based domain with abstract values
    $\{\top,\text{pos}_x,\text{neg}_x,x\!>\!\!10,x\!\leq\!\!10,\bot\}$.
\par\vskip -2ex {\small 
   \begin{align*}
        p(x) \clauseif       
\anow{x\dot{>}0&}{\ahiding{x'}{(\aparallel{\aparallel{\atell{x=[\_|x']}}{\atell{
x'=[x+1|\_]}}
        }{p(x') } )} \\
&\;\,}{\ahiding{x''}{(\aparallel{\aparallel{\atell{x=[\_|x'']}}{\atell{x''=[
x-1|\_]}}
        }{p(x'')}
        })}
    \end{align*} 
}%
    Due to the monotonicity of the store, we have to use streams (written in a list-fashion way) to model the \emph{imperative-style} variables \cite{deBoerGM99}.
In this way, variable $x$ in the program above is a stream that is updated with different values during the execution.
    Following this idea, the abstraction for concrete streams is defined as the
(abstracted) last instantiated value in the stream.
    The concretization of one stream is defined as all the concrete streams whose last value is a concretization of the abstract one.
We write a dot on a predicate symbol (\eg\ $\dot{=}$) to denote that we want to check it for the last instantiated value of a stream.

    We define the following intended specification to specify that, (a) if the parameter 
    is greater than 10, then the last value of the stream (written $\dot{x}$) will always
    be greater than 10; (b) if the parameter is negative, then the value is always negative
\par\vskip -3ex {\small
    \begin{align*}
        \Sz(p(x_1) = %&
        \{\Acrs[+\infty]{({x}_1\mathrm{\dot{>}\!10},\ACSfalseDual)}
        {{x}\mathrm{\dot{>}\!10}}
        {{x}\mathrm{\dot{>}\!10}}\}%\\
        %&
\cup\{ \Acrs[+\infty]{(\mathrm{neg}_{\dot{x}},\ACSfalseDual)} {\mathrm{neg}_{\dot{x}}}
        {\mathrm{neg}_{\dot{x}}} \}
    \end{align*}
}%
The two abstract sequences represent infinite computations thanks to the
$+\infty$ index in the last tuple.
In other words, finite specifications that represent 
infinite computations can be considered and effectively handled.
In fact, we can compute $\Tp{\{d\}}{\Sz}$:
\par\vskip -2ex {\small
    \begin{align*}
        &\{
        \begin{aligned}[t]
            &           
            \{\Acrs[1]{(\mathrm{neg}_{\dot{x}},\ACSfalseDual)}
            {\mathrm{pos}_{\dot{x}}}
            {\mathrm{pos}_{\dot{x}}}\cdot
            (\Aprop{\mathrm{pos}_{\dot{x}}}{\Acrs[+\infty]{(\dot{x}\mathrm{>\!\!10},\ACSfalseDual)}
            {\dot{x}\mathrm{>\!\!10}}
            {\dot{x}\mathrm{>\!\!10}}})\}\\           
            &\cup\{
            \Acrs[1]{(\mathrm{neg}_{\dot{x}},\ACSfalseDual)} {\mathrm{neg}_{\dot{x}}}
            {\mathrm{neg}_{\dot{x}}}\cdot
            \Acrs[+\infty]{(\mathrm{neg}_{\dot{x}},\ACSfalseDual)} {\mathrm{neg}_{\dot{x}}}
            {\mathrm{neg}_{\dot{x}}}
            \}\}
        \end{aligned}\\[-1ex]
&= \\[-3ex]
        &\{
        \begin{aligned}[t]
            &           
            \{\Acrs[1]{(\mathrm{neg}_{\dot{x}},\ACSfalseDual)}
            {\mathrm{pos}_{\dot{x}}}
            {\mathrm{pos}_{\dot{x}}}\cdot
            \Acrs[+\infty]{(\overbrace{\mathrm{pos}_{\dot{x}}\ACSmerge\dot{x}\mathrm{>\!\!10}}^{\mathrm{pos}_{\dot{x}}},\ACSfalseDual)}
            {\mathrm{pos}_{\dot{x}}\ACSmerge\dot{x}\mathrm{>\!\!10}}
            {\mathrm{pos}_{\dot{x}}\ACSmerge\dot{x}\mathrm{>\!\!10}})\}\\           
            &\cup\{
            \Acrs[1]{(\mathrm{neg}_{\dot{x}},\ACSfalseDual)} {\mathrm{neg}_{\dot{x}}}
            {\mathrm{neg}_{\dot{x}}}\cdot
            \Acrs[+\infty]{(\mathrm{neg}_{\dot{x}},\ACSfalseDual)} {\mathrm{neg}_{\dot{x}}}
            {\mathrm{neg}_{\dot{x}}}
            \}\}
        \end{aligned}\\[-1ex]
&= \\[-1ex]
&\{\begin{aligned}[t]            
\{\Acrs[+\infty]{(\mathrm{pos}_{\dot{x}},\ACSfalseDual)}
            {\mathrm{pos}_{\dot{x}}}
            {\mathrm{pos}_{\dot{x}}}\} %\\           
            \cup
\{
            \Acrs[+\infty]{(\mathrm{neg}_{\dot{x}},\ACSfalseDual)} {\mathrm{neg}_{\dot{x}}}
            {\mathrm{neg}_{\dot{x}}}  
            \}\}
\end{aligned}
    \end{align*}
}%
The third equality holds because $\mathrm{pos}_{\dot{x}}$ entails ${x}\mathrm{\dot{>}\!10}$,
so the merge of the two constraints will be equal to $\mathrm{pos}_{\dot{x}}$.

Since $\Tp{\{d\}}{\Sz}\not\leqA \Sz$ we can conclude that $d$ is an incorrect declaration \wrt{} $\Sz$.
In addition, we can notice that $\Sz$ contain an uncovered
element that is a sequence that cannot be derived by the semantics operator.
\end{example}

\section{Related Work}
A top-down (big-step) denotational semantics for \tccp{} is defined in
\cite{deBoerGM99} for terminating computations. 
In that work, a terminating computation is both, a computation that 
reaches a point in which no agents are pending to be executed, and also 
a computation that suspends since there is no enough information in the 
store to make the choice agents evolve.
Our semantics is a bottom-up (small-step) denotational 
semantics that models infinite computations, and also distinguishes the two 
kinds of terminating computations aforementioned. Conceptually, a suspended 
computation has not completely finished
its execution, and, in some cases, it could be a symptom of a system error.
Thus, the new semantics is well suited to handle, not only functional 
systems (where an input-output semantics makes sense), but also 
reactive systems. 

In \cite{FalaschiOPV07}, a first approach to the declarative
debugging of a \ccp{} language is presented.  
However, it does not cover the particular
extra difficulty of the non-monotonicity, common to all timed concurrent 
constraint languages.
As we have said, this ability is crucial in order to
model specific behaviors of reactive systems, such as timeouts or
preemption actions.
This is the main reason why our abstract (and concrete) semantics are significantly
different from \cite{FalaschiOPV07} and from formalizations for other
declarative languages.

The idea of using two different mechanisms for dealing with positive and 
negative information in our abstraction scheme is inspired by \cite{AlpuenteGPV05}.
There, a framework for the abstract model checking of \tccp{} programs
based on a source-to-source transformation is defined. In particular, it is
defined
a transformation from a
\tccp{}
program $P$ into a \tccp{} program $\bar{P}$ that represents a correct abstraction 
of the original one (in the sense that the semantics of $P$ are
included in the
semantics of $\bar{P}$). % is defined. 
Instead, we define an abstract semantics
for the language. The upper- and lower-approximated versions of the 
entailment relation are used in order to keep $\bar{P}$ correct, but also precise enough.

\section{Conclusion and Future Work}

We have presented a new compact, bottom-up semantics for the \tccp{}
language which is correct and fully abstract w.r.t. the behavior of the
language.  This semantics is well suited for debugging and verification
purposes in the context of reactive systems.
The idea of using conditions in order to have a correct bottom-up semantics 
can be also applied to other non-monotonic languages such as, for
example, \textsf{ntcc} in the \ccp{} paradigm \cite{PalamidessiV-CP2001} or
\textsf{Linda} in the imperative (coordination) paradigm
\cite{Gelernter85}. 

Then, an abstract semantics that is able to specify (a kind of) infinite 
computations is presented. It is based on the abstraction of computation sequences
by using two functions that satisfy some properties in order to guarantee correctness. 
All our examples satisfy those conditions.
The abstract semantics keeps the synchronization among
parallel computations, which is a particular difficulty of the \tccp{} language.
As already noticed in \cite{AlpuenteGPV05}, the loss of synchronization in other 
\ccp{} languages just implies a loss of precision, but in the case of \tccp{}, due 
to the maximal parallelism, it would imply a loss of correctness.

Finally, we have adapted the abstract diagnosis approach to the \tccp{}
language employing the new semantics as basis.  We have presented two
illustrative examples to show the new features of our approach w.r.t.
other paradigms. %, and also its limitations.  %\Alicia{maybe too ``hard''}

As future work, we intend to work on abstractions of our semantics to
domains of temporal logic formulas, in order to be able to specify safety
and/or liveness properties, and to compare its models w.r.t. the program
semantics. Another interesting aspect is to study if a general framework 
for the proposed methodology can be defined in order to apply it to other
languages.

\bibliographystyle{acmtrans}%unsrt
\bibliography{biblioICLP}%

\end{document}